# QUANTIFYING NONORTHOGONALITY

Oliver Cohen

Department of Physics, Carnegie Mellon University, Pittsburgh, PA 15213, and Theoretical Physics Research Unit, Birkbeck College, University of London, Malet Street, London WC1E 7HX, UK

e-mail: o.cohen@andrew.cmu.edu

Abstract

An exploratory approach to the possibility of analyzing nonorthogonality as a quantifiable property is presented. Three different measures for the nonorthogonality of pure states are introduced, and one of these measures is extended to single-particle density matrices using methods that are similar to recently introduced techniques for quantifying entanglement. Several interesting special cases are considered. It is pointed out that a measure of nonorthogonality can meaningfully be associated with a single mixed quantum state. It is then shown how nonorthogonality can be unlocked with classical information; this analysis reveals interesting inequalities and points to a number of connections between nonorthogonality and entanglement.

Quantum information theory has made rapid progress in the last few years, and has led to the introduction of a number of novel information processing techniques [1-4]. Several [2-5] of these techniques depend on the use of *entangled* quantum systems for their efficacy, and consequently entanglement has come to be seen as a valuable resource, necessary for the implementation of such techniques. Quantification of entanglement for the full range of quantum states is a difficult problem on which much work has been, and continues to be, carried out [6-9].

However, for quantum cryptography, which was the first of the new methods of quantum information processing to be discovered [10], the use of entangled states is not essential. Although it is possible to devise quantum cryptography schemes that do use entanglement [5, 11, 12], the earliest and best known schemes [1, 10], which have also been the most effective in laboratory implementations, do not involve the use of any entangled states. Nevertheless, these schemes do require the use of *nonorthogonal* states. If the states used in these schemes were restricted to mutually orthogonal ones, then, although it would still be possible to encode cryptographic key data using these states, passive eavesdropping could not be ruled out. Hence the exclusion of nonorthogonal states would effectively render these schemes redundant, since their crucial advantage over corresponding classical schemes is their ability to eliminate the possibility of passive eavesdropping. Moreover, any two nonorthogonal states are, by themselves, sufficient to implement a quantum cryptography scheme [13].

So it seems reasonable to consider the possibility that nonorthogonality could be viewed as a resource in an analogous way to that in which we now think of entanglement. One might object to this characterization and point out that, given a set of quantum systems, we can always prepare them in nonorthogonal states if we wish. However, a similar argument would apply to entanglement: given a collection of bipartite quantum systems, we can always in principle prepare them in entangled states by carrying out Bell operator basis [14] measurements. We propose, then, that nonorthogonality be considered as a potential resource; it is certainly essential for quantum cryptography schemes such as the well known "BB84" [1] method. Like entanglement, nonorthogonality has no classical equivalent. In this paper we will examine various ways in which nonorthogonality can be quantified.

The two properties of entanglement and nonorthogonality are embodied by quantum systems in characteristically different ways. Whereas we can meaningfully discuss the entanglement of a *single* quantum state, this does not normally apply to nonorthogonality, which, in the case of pure states, necessarily refers to at least two states. This discrepancy appears less acute, however, when we consider that an entangled state must refer to at least two subsystems, to each of which can be assigned an individual mixed state (or a pure state if the entangled state becomes disentangled); thus in a sense an entangled state also subsumes at least two separate individual states, albeit that these states refer to different (sub)systems.

In this paper we consider a number of possible measures for nonorthogonality that are applicable to pairs of pure states. We focus in particular on a simple linear measure. We explain how this measure can be extended to general density matrices by using a method similar to that employed in analyzing hidden entanglement [8] and entanglement of assistance [9]. We show how nonorthogonality can be unlocked with classical information in an analogous way to that in which hidden entanglement can be unlocked. This analysis reveals interesting inequalities for unlocked nonorthogonality, which parallel previously proposed inequalities for unlocked entanglement, and point to further connections between nonorthogonality and entanglement.

There are a number of ways in which one might begin to approach the question of how nonorthogonality could be quantified. A particularly simple approach would be to adopt a linear measure, as follows. Suppose we are given two quantum states, $|\psi_1\rangle$ and $|\psi_2\rangle$. We could then introduce the measure $N_0$ such that the amount of nonorthogonality associated with this pair of states is given by:

$$N_0(|\psi_1\rangle, |\psi_2\rangle) = 1 - 2\left||\langle\psi_2|\psi_1\rangle|^2 - 1/2\right|. \tag{1}$$

Adopting this measure means that two identical states (or two states differing only by a phase factor) will have zero nonorthogonality, as will two orthogonal states. (Attributing zero nonorthogonality to two identical states may seem contradictory, but the relevant point is that two such states can obviously be included in a single orthogonal set.) The maximum possible nonorthogonality, 1, occurs when $|\langle\psi_2|\psi_1\rangle|^2 = 1/2$. Intermediate cases yield values of $N_0$ that reflect the probability of predicting the wrong result if a quantum system in state $|\psi_2\rangle$ is measured using an operator of which $|\psi_1\rangle$ is an eigenstate (or vice-versa).

Thus $N_0$ has a clear physical meaning, in correspondence with the probability of producing an error when we try to determine one of the states $|\psi_1\rangle, |\psi_2\rangle$ using a measurement basis encompassing the other state. (Actually $N_0$ is chosen to be twice this probability.) To see how it might be used, consider a generalization of the BB84 [1] quantum cryptography scheme, where, rather than necessarily using two *maximally* nonorthogonal pairs of orthogonal states, we use two pairs of nonorthogonal states where the nonorthogonality between each pair is arbitrary. That is, we use two nonorthogonal pairs of orthogonal two-dimensional states $(|\alpha_\uparrow\rangle, |\alpha_\downarrow\rangle)$ and $(|\beta_\uparrow\rangle, |\beta_\downarrow\rangle)$, where $\langle \alpha_\uparrow | \alpha_\downarrow \rangle = \langle \beta_\uparrow | \beta_\downarrow \rangle = 0$ and $0 < |\langle \alpha_\uparrow | \beta_\uparrow \rangle|^2 < 1$, but with $|\langle \alpha_\uparrow | \beta_\uparrow \rangle|^2$ not necessarily equal to ½ as it is in the standard BB84 scheme. Then it can easily be shown that, if the amount of nonorthogonality associated with the pair of states $(|\alpha_\uparrow\rangle, |\beta_\uparrow\rangle)$ is $\tilde{N}_0$, then the nonorthogonality associated with each of the pairs $(|\alpha_\uparrow\rangle, |\beta_\downarrow\rangle)$, $(|\alpha_\downarrow\rangle, |\beta_\uparrow\rangle)$, and $(|\alpha_\downarrow\rangle, |\beta_\downarrow\rangle)$ must also be $\tilde{N}_0$. We can also express the probability of detecting an eavesdropper ("Eve") in terms of $\tilde{N}_0$. Whereas in the standard BB84 scheme the probability of detecting Eve in a single transmission in which sender Alice and receiver Bob have used the same measurement basis is ¼, in our more general scheme we find that the detection probability is $\frac{\tilde{N}_0}{2}\left(1 - \frac{\tilde{N}_0}{2}\right)$. (We have assumed in both cases that Eve simply chooses randomly between the two possible bases and carries out an ideal measurement.)

A similar result can be obtained for the BB92 scheme [13] which uses only two transmission states $|u_0\rangle$ and $|u_1\rangle$, which are necessarily nonorthogonal. In this scheme Bob uses two noncommuting projection operators, $P_0$ and $P_1$, where $P_0 = 1 - |u_1\rangle\langle u_1|$ and $P_1 = 1 - |u_0\rangle\langle u_0|$. If $|u_0\rangle$ and $|u_1\rangle$ are both two-dimensional and Eve chooses randomly between measurement bases encompassing $|u_0\rangle$ and $|u_1\rangle$, then the probability of detecting her for a single transmission in which Alice sends $|u_0\rangle$

($|u_1\rangle$) and Bob measures $P_0$ ($P_1$) will be $\frac{\bar{N}_0}{2}\left(1-\frac{\bar{N}_0}{2}\right)$, where $\bar{N}_0$ is the nonorthogonality associated with $|u_0\rangle$ and $|u_1\rangle$. If instead Eve chooses randomly between $P_0$ and $P_1$, the single-transmission detection probability reduces to $\frac{\bar{N}_0}{4}\left(1-\frac{\bar{N}_0}{2}\right)$.

Whilst $N_0$ has a straightforward physical interpretation, it does not have the kind of quasi-entropic form that is familiar from classical information theory and from the standard measure for quantifying pure state entanglement. A second measure for the nonorthogonality of two states $|\psi_1\rangle$, $|\psi_2\rangle$, which is more closely related to the standard entanglement measure is given by $N_1$, where

$$N_1(|\psi_1\rangle,|\psi_2\rangle) = -\left\{|\langle\psi_2|\psi_1\rangle|^2 \log_2 |\langle\psi_2|\psi_1\rangle|^2 + \left(1-|\langle\psi_2|\psi_1\rangle|^2\right)\log_2\left(1-|\langle\psi_2|\psi_1\rangle|^2\right)\right\}. \quad (2)$$

$N_1$ can also be given a physical interpretation as follows. It represents the amount of *absolutely selective* [15] classical information (measured in bits) that is generated by carrying out a measurement on $|\psi_2\rangle$ using a two-dimensional measurement basis encompassing $|\psi_1\rangle$, or vice-versa. By "absolutely selective" information we mean the number of bits of genuinely *new* data that is generated through the measurement process. It should be emphasised that this type of information does not arise merely from the removal of ignorance on the part of the experimenter, but represents fundamentally unpredictable data, and hence can justifiably be described as absolutely selective. This type of information does not arise in a classical deterministic theory and can be seen as a basic distinguishing feature with respect to quantum and classical physics.

The standard measure of entanglement for pure states of two subsystems can, like $N_1$, also be equated to a quantity of absolutely selective classical information, specifically the amount of data that is generated by measuring one of the subsystems in the Schmidt basis. Thus if we have an entangled state $|\psi_{12}\rangle$ with biorthogonal decomposition $\sum_i \alpha_i |\phi_{1i}\rangle |\chi_{2i}\rangle$, the entanglement of $|\psi_{12}\rangle$ is $\xi$ ebits, where $\xi = -\sum_i |\alpha_i|^2 \log_2 |\alpha_i|^2$. But this quantity is also equal to the amount of absolutely selective classical information, measured in bits, that can be generated by measuring either of the subsystems in the basis given by the Schmidt decomposition.

Interestingly, the quantity $\xi$ can also be identified as the *minimum* amount of absolutely selective information that can be generated by measuring one of the subsystems of the state $|\psi_{12}\rangle$, given a free choice of measurement basis. It can easily be shown that, if we choose any measurement basis for measuring one of the subsystems other than that given by the Schmidt decomposition, the amount of genuinely new data generated will be greater than $\xi$.

By analogy with this, we can introduce a third measure of nonorthogonality $N_2$, which is given by the minimum amount of absolutely selective information that can be generated by carrying out measurements on two nonorthogonal states $|\psi_1\rangle$, $|\psi_2\rangle$ using only a single measurement basis. In other words $N_2$ would be determined by selecting the measurement basis that would minimize the new data generated by performing measurements on $|\psi_1\rangle$ and $|\psi_2\rangle$. Both $N_1$ and $N_2$ provide links between classical information and nonorthogonality, which, like entanglement, can be thought of as a wholly quantum form of information.

In this paper we focus on the first and simplest of the proposed measures, the linear measure $N_0$ (which from now on we will denote simply by *N)*. This measure leads to interesting results when it is applied to states that are represented by arbitrary density matrices. Suppose for example we consider the diagonalized density matrix $\rho$, with

$$\rho = p|\uparrow\rangle\langle\uparrow| + (1-p)|\downarrow\rangle\langle\downarrow|, \tag{3}$$

where $|\uparrow\rangle$ and $|\downarrow\rangle$ are orthogonal states of a two-state quantum system.

As is generally the case for density matrices, there are two physically distinct scenarios that can both be described by $\rho$. On the one hand, $\rho$ could represent an ensemble of quantum systems, where each individual system has been prepared in a definite quantum state but where we are ignorant as regards exactly which state each system is in. (In the language of [8], such an ensemble could be described as being in a *pseudomixed* state). Alternatively, $\rho$ could be derived from an entangled two-system state such as $|\psi_{12}\rangle = \sqrt{p}|\uparrow_1\uparrow_2\rangle + \sqrt{1-p}|\downarrow_1\downarrow_2\rangle$, with system 2 traced out.

In the former case we can introduce the notion of *hidden* nonorthogonality, by considering different decompositions of $\rho$ in terms of nonorthogonal states. If we are told that an ensemble described by $\rho$ has in fact been prepared using a specific decomposition of nonorthogonal states, we can then *unlock* the nonorthogonality, without performing any measurements, if we are supplied with the relevant classical information to establish which state each system is in.

In the latter case, where $\rho$ is derived from an entangled state such as $|\psi_{12}\rangle$ above, we can consider different ways in which we can *prepare* nonorthogonal states of system 1 by carrying out measurements in suitable bases on system 2. We could then, for example, examine how we can maximize the nonorthogonality associated with the prepared states of system 1 by selecting the appropriate measurement basis for system 2. The nonorthogonality obtained in this way could be described as "nonorthogonality of assistance".

These two approaches correspond to the techniques relating to hidden entanglement [8] and entanglement of assistance [9] respectively. It is worth pointing out that entanglement of assistance was originally conceived as dual to the entanglement of formation [7], which is a measure of the *minimum* amount of entanglement required to prepare a given bipartite density matrix. However, the corresponding "nonorthogonality of formation" for a single-particle density matrix will always be zero, since the density matrix can always be decomposed exclusively in terms of orthogonal states.

Our chosen approach in this paper will focus on hidden nonorthogonality, which will facilitate the straightforward derivation of several interesting results. Suppose, then, that our density matrix $\rho$, given by eq. (3), in fact represents an ensemble of systems, each of which has been prepared in one or other of two nonorthogonal states. We describe these states as *hidden* nonorthogonal states, because we will not be able to determine with certainty which state each system has been prepared in unless we are supplied with the relevant classical information pertaining to the preparation. Indeed, without this information we will be unable to make use of the nonorthogonality associated with the preparation states.

By using a method similar to that used in [8], it can easily be shown that there are infinite ways of decomposing $\rho$ in terms of two nonorthogonal states, $|\phi_1\rangle$ and $|\phi_2\rangle$. Such a decomposition can be written as:

$$\rho = z|\phi_1\rangle\langle\phi_1| + (1-z)|\phi_2\rangle\langle\phi_2|, \tag{4}$$

where
$$|\phi_1\rangle = z^{-1/2}\left(\sqrt{p}\alpha|\uparrow\rangle + \sqrt{1-p}\beta^*|\downarrow\rangle\right) \tag{5a}$$

and
$$|\phi_2\rangle = (1-z)^{-1/2}\left(\sqrt{p}\beta|\uparrow\rangle - \sqrt{1-p}\alpha^*|\downarrow\rangle\right). \tag{5b}$$

In eq.s (5) $\alpha$ and $\beta$ can be any two complex numbers satisfying $|\alpha|^2 + |\beta|^2 = 1$; and $z = 2p|\alpha|^2 + 1 - p - |\alpha|^2$.

We can thus think of $|\phi_1\rangle$ and $|\phi_2\rangle$ as the hidden nonorthogonal states used in the preparation of $\rho$. It follows from eq.s (5) that

$$|\langle\phi_2|\phi_1\rangle|^2 = 1 - \frac{p(1-p)}{z(1-z)}, \tag{6}$$

and the amount $N_{\phi_1\phi_2}$ of nonorthogonality associated with the pair of states $|\phi_1\rangle$ and $|\phi_2\rangle$ is given by

$$N_{\phi_1\phi_2} = 1 - 2\left|\frac{1}{2} - \frac{p(1-p)}{z(1-z)}\right|. \tag{7}$$

We can also introduce a measure of nonorthogonality that refers to the whole ensemble of systems, rather than to the two states represented in the ensemble. This measure we denote $(N_{ens.})_{\phi_1\phi_2}$, where

$$(N_{ens.})_{\phi_1\phi_2} = 2\min(z,(1-z))N_{\phi_1\phi_2}. \tag{8}$$

The factor $2\min(z,(1-z))$ in eq. (8) reflects the fact that we will be able to pair off into nonorthogonal states only a fraction $2\min(z,(1-z))$ of the whole ensemble of systems. Thus $(N_{ens.})_{\phi_1\phi_2}$ can be interpreted as the average amount of nonorthogonality per pair of systems in the ensemble. Without loss of generality we can take $p \geq 1/2$, and $|\alpha|^2 \geq 1/2$, from which it follows that $z \geq 1/2$ so that $\min(z,(1-z)) = 1-z$. We then find that, for $|\langle\phi_2|\phi_1\rangle|^2 < 1/2$,

$$(N_{ens.})_{\phi_1\phi_2} = 4(1-z) - \frac{4p(1-p)}{z} \tag{9a}$$

whilst for $|\langle\phi_2|\phi_1\rangle|^2 \geq 1/2$

$$(N_{ens.})_{\phi_1\phi_2} = \frac{4p(1-p)}{z}. \tag{9b}$$

We now consider some interesting special cases. First of all, we can identify those two-state preparations of a given density matrix that will yield maximum nonorthogonality between these states. The condition for this is $N_{\phi_1\phi_2} = 1$, from which we obtain $z(1-z) = 2p(1-p)$. Recalling that we are assuming that $z \geq 1/2$, it follows that the unique permissible value of $z$ satisfying this equation is given by $z = 1/2(1+\sqrt{1-8p(1-p)})$. However this value of $z$ can be real only if either $p \geq 1/2(1+\sqrt{2}/2)$ or $p \leq 1/2(1-\sqrt{2}/2)$. Hence, since we are assuming that $p \geq 1/2$, this shows that we can obtain maximum nonorthogonality between two states in an ensemble described by $\rho$ only if $p \geq 1/2(1+\sqrt{2}/2)$. For density matrices satisfying this inequality, we will obtain maximum nonorthogonality between the preparation states when $z = 1/2(1+\sqrt{1-8p(1-p)})$, or, equivalently, when

$$|\alpha|^2 = 1/2\left(\frac{\sqrt{1-8p(1-p)}}{2p-1}\right).$$

Rather than looking at nonorthogonality between individual states in an ensemble described by $\rho$, we can also examine how we can obtain the maximum yield of nonorthogonality for the *whole* ensemble, taking into account the fact that it will not in general be possible to pair off into nonorthogonal states all the systems in an ensemble. From eq.s (9) we can see that maximum nonorthogonality for the whole ensemble is obtained when $(1-z) - \frac{p(1-p)}{z}$ is a maximum, if $|\langle\phi_2|\phi_1\rangle|^2 < 1/2$, and when $\frac{p(1-p)}{z}$ is a maximum, if $|\langle\phi_2|\phi_1\rangle|^2 \geq 1/2$. We find that, in both these cases, maximum nonorthogonality for the ensemble is obtained when $z = 1/2$, and that,

for $|\langle\phi_2|\phi_1\rangle|^2 < 1/2$, $\quad [(N_{ens.})_{\phi_1\phi_2}]_{max.} = 2 - 8p(1-p)$ (10a)

whilst for $|\langle\phi_2|\phi_1\rangle|^2 \geq 1/2,$ $\quad [(N_{ens.})_{\phi_1\phi_2}]_{max.} = 8p(1-p).$ (10b)

Since $[(N_{ens.})_{\phi_1\phi_2}]_{max.}$ is uniquely determined for a given value of $p$, we can see that for any density matrix $\rho$ of a single two-dimensional system as given by eq. (3), the quantity $[(N_{ens.})_{\phi_1\phi_2}]_{max.}$ is well defined. Hence it is quite meaningful to associate the quantity $[(N_{ens.})_{\phi_1\phi_2}]_{max.}$ with a *single* mixed quantum state $\rho$, which in turn can be associated with a single quantum system. So the measure $[(N_{ens.})_{\phi_1\phi_2}]_{max.}$ demonstrates that nonorthogonality is a property that can meaningfully be defined for a single mixed quantum state and indeed for a single system that is in such a state.

It is possible to combine the results obtained earlier in order to determine the "ideal" case, where we obtain maxima for both $N_{\phi_1\phi_2}$ and $(N_{ens.})_{\phi_1\phi_2}$. The conditions for this are (i) $z = 1/2(1+\sqrt{1-8p(1-p)})$, and (ii) $z=1/2$. Hence our ideal case is given by $p = 1/2(1+\sqrt{2}/2)$, which yields the density matrix $\tilde{\rho}$, where

$$\tilde{\rho} = 1/2(1+\sqrt{2}/2)|\uparrow\rangle\langle\uparrow| + 1/2(1-\sqrt{2}/2)|\downarrow\rangle\langle\downarrow|.$$ (11)

In other words, the density matrix $\tilde{\rho}$, and only this density matrix, can represent a 50/50 mixture of maximally nonorthogonal states.

We now assess how much classical information is needed to *unlock* the nonorthogonality that may be hidden in an ensemble described by a density matrix such as $\rho$ given by eq. (3). In other words, we want to determine how much classical information is needed to distinguish, without performing any measurements, the different states in the ensemble. If the hidden nonorthogonal states are $|\phi_1\rangle$ and $|\phi_2\rangle$ given by eq.s (5), then, in order to unlock all the nonorthogonality associated with the ensemble, we will require sufficient information to identify which individual systems are in each state. The amount of information needed to do this is $U$ bits per system, where

$$U = -[z \log_2 z + (1-z)\log_2(1-z)]. \tag{12}$$

It is worth noting that $U$ is always greater than or equal to $I$, where $I = -[p \log_2 p + (1-p)\log_2(1-p)]$ is the number of bits of classical information per system necessary to distinguish the states in an ensemble described by $\rho$ when it is prepared with *orthogonal* states. (This inequality follows from the fact that, as can easily be shown, $z$ is always less than or equal to $p$.) It is thus possible to identify a quantity $E = U - I$, which represents, for a given density matrix $\rho$, the *excess* classical information per system needed to identify the states in the ensemble when it is prepared using a pair of nonorthogonal as opposed to orthogonal states. In other words, $E$ represents the difference between the respective amounts of classical information needed to identify, without performing any measurements, the states in nonorthogonal and orthogonal preparations of an ensemble described by a given density matrix. The quantity $E$ can be interpreted as the extra price, in classical information, that we have to pay in order to prepare a given density matrix with nonorthogonal states and then identify these states, as opposed to carrying out the corresponding process with orthogonal states. This interpretation of $E$ provides another link between classical information and nonorthogonality.

Next we assess the quantity $\frac{2U}{(N_{ens.})_{\phi_1\phi_2}}$, which represents the number of bits of classical information required to unlock each unit of nonorthogonality (or "nbit") contained in the ensemble. The factor 2 occurs here because $(N_{ens.})_{\phi_1\phi_2}$ refers to the average amount of nonorthogonality per *pair* of systems in the ensemble, whereas $U$ refers to the amount of classical information needed per *individual* system in order to distinguish the states. Analysis of a large number of examples indicates that $\frac{U}{(N_{ens.})_{\phi_1\phi_2}}$ is always greater than or equal to 1. In other words it seems that we always need at least two bits to unlock each nbit. This closely parallels a result in [8], where it was conjectured that we always need at least one bit to unlock each ebit.

By contrast we find that $\frac{E}{(N_{ens.})_{\phi_1\phi_2}}$ is usually (but not always) less than 1. This means that the *excess* classical information needed to unlock each nbit is usually less than 2 bits.

This Letter is intended as an initial exploratory approach to the analysis of nonorthogonality as a quantifiable property, and consequently only the most elementary cases have been considered. In particular, our analysis has been restricted to cases where only two distinct nonorthogonal states are involved. Clearly, however, it is possible to prepare an ensemble using any number of distinct nonorthogonal states. The quantification of nonorthogonality for such an ensemble will be considerably more complex, as there will no longer be a unique way of pairing off the nonorthogonal states, and so a prescription for this pairing will be necessary.

To conclude, we have seen that nonorthogonality can be analyzed and quantified using techniques that are similar to some of those that have been used to analyze and quantify entanglement. Nonorthogonality can thus be understood as a separate, independent, quantifiable property for applications in quantum information theory.

I am grateful to Todd Brun and Robert Griffiths for interesting and helpful discussions. This research was supported by the Leverhulme Trust and by NSF Grant No. PHY-9602084.